\documentclass[preprint2]{aastex}

\shorttitle{LATITUDE DEPENDENCE OF DIFFUSE GALACTIC LIGHT}

\shortauthors{Sano et al.}

\begin{document}


\title{GALACTIC LATITUDE DEPENDENCE OF NEAR-INFRARED DIFFUSE GALACTIC LIGHT : THERMAL EMISSION OR SCATTERED LIGHT?}


\author{K. SANO\altaffilmark{1,2,3} and
 %K. KAWARA\altaffilmark{3}, 
 S. MATSUURA\altaffilmark{3}
 %K. TSUMURA\altaffilmark{4}}
 %H. KATAZA\altaffilmark{1,2}, 
%T. ARAI\altaffilmark{4}, 
%M. SHIRAHATA\altaffilmark{4}, 
%Y. ONISHI\altaffilmark{2,5}}
%Y. MATSUOKA\altaffilmark{6}
}

\affil{
\altaffilmark{1}Department of Astronomy, Graduate School of Science, The University of Tokyo, \\
Hongo 7-3-1, Bunkyo-ku, Tokyo 113-0033, Japan\\
\altaffilmark{2}Institute of Space and Astronautical Science, Japan Aerospace Exploration Agency,\\
3-1-1 Yoshinodai, Chuo-ku, Sagamihara, Kanagawa 252-5210, Japan\\
\altaffilmark{3}Department of Physics, School of Science and Engineering, Kwansei Gakuin University, 2-1 Gakuen, Sanda, Hyogo 669-1337, Japan \\
}

\email{sano0410@kwansei.ac.jp}

%\and

%\author{R. J. Hanisch\altaffilmark{5}}
%\affil{Space Telescope Science Institute, Baltimore, MD 21218}


%\altaffiltext{1}{Visiting Astronomer, Cerro Tololo Inter-American Observatory.
%CTIO is operated by AURA, Inc.\ under contract to the National Science
%\altaffiltext{2}{Society of Fellows, Harvard University.}
%\altaffiltext{3}{present address: Center for Astrophysics,
 %   60 Garden Street, Cambridge, MA 02138}
%\altaffiltext{4}{Visiting Programmer, Space Telescope Science Institute}
%\altaffiltext{5}{Patron, Alonso's Bar and Grill}


\begin{abstract}
Near-infrared (IR) diffuse Galactic light (DGL) consists of scattered light and thermal emission from interstellar dust grains illuminated by interstellar radiation field (ISRF).
At $1.25$ and $2.2\,\rm{\mu m}$, recent observational study shows that intensity ratios of the DGL to interstellar $100\,\rm{\mu m}$ dust emission steeply decrease toward high Galactic latitudes ($b$).
In this paper, we investigate origin(s) of the $b$-dependence on the basis of models of thermal emission and scattered light. 
Combining a thermal emission model with regional variation of the polycyclic aromatic hydrocarbon abundance observed with {\it Planck}, we show that contribution of the near-IR thermal emission component to the observed DGL is less than $\sim 20\%$.
We also examine the $b$-dependence of the scattered light, assuming a plane-parallel Galaxy with smooth distributions of the ISRF and dust density along vertical direction, and assuming a scattering phase function according to a recently developed model of interstellar dust.
We normalize the scattered light intensity to the $100\,\rm{\mu m}$ intensity corrected for deviation from the cosecant-$b$ law according to the {\it Planck} observation. 
As the result, the present model taking all the $b$-dependence of dust and ISRF properties can account for the observed $b$-dependence of the near-IR DGL.
However, uncertainty of the correction for the $100\,\rm{\mu m}$ emission is large and other normalizing quantities may be appropriate for more robust analysis of the DGL.
\end{abstract}





\section{INTRODUCTION}

Dust is a minor constituent by mass in the universe, but plays crucial roles in various astrophysical phenomena.
As a basic property of dust grains, they scatter starlight and absorb the radiation energy.
Subsequently, they release the absorbed energy as thermal emission in infrared (IR) wavelengths.
In the diffuse interstellar medium (ISM), the scattered light component is dominant from ultraviolet (UV) to near-IR wavelengths ($\sim0.2$--$2\,\rm{\mu m}$).
From near to mid-IR ($\sim2$--$50\,\rm{\mu m}$), very small grains and large molecules (e.g., polycyclic aromatic hydrocarbon; PAH) heated by interstellar radiation field (ISRF) radiate thermal emission.
From UV to near-IR, the scattered light and thermal emission in the diffuse ISM are conventionally referred to as ``diffuse Galactic light (DGL)''. 

The DGL observation is useful to investigate interstellar dust properties such as grain albedo and the PAH abundance.
As a tracer of the DGL, far-IR $100\,\rm{\mu m}$ dust emission has been used, because they are expected to correlate linearly with each other in optically thin fields (e.g., Brandt \& Draine 2012; hereafter BD12).
The $100\,\rm{\mu m}$ map based on all-sky observations of {\it Infrared Astronomical Satellite} ({\it IRAS}) and {\it Cosmic Background Explorer} ({\it COBE}) has been used frequently (Schlegel et al. 1998; hereafter SFD98).
From UV to near-IR, several studies have observed the diffuse light in high Galactic latitudes ($b$) and found linear correlations against the $100\,\rm{\mu m}$ emission (e.g., Witt et al. 2008; Matsuoka et al. 2011; Tsumura et al. 2013; Arai et al. 2015; Sano et al. 2015; 2016a; Kawara et al. 2017).
These results are shown in Figure 1 with model spectra of the scattered light (BD12) and thermal emission (Draine \& Li 2007; hereafter DL07).
The models are based on recent interstellar dust models developed by Weingartner \& Draine (2001; hereafter WD01) and Zubko et al. (2004), assuming the ISRF spectrum in the solar neighborhood (Mathis et al. 1983; hereafter MMP83).
As shown in Figure 1, the near-IR DGL is expected to consist of the scattered light and thermal emission, indicating that both components should be taken into account in detailed study.

\begin{figure*}
\begin{center}
 \includegraphics[scale=0.8]{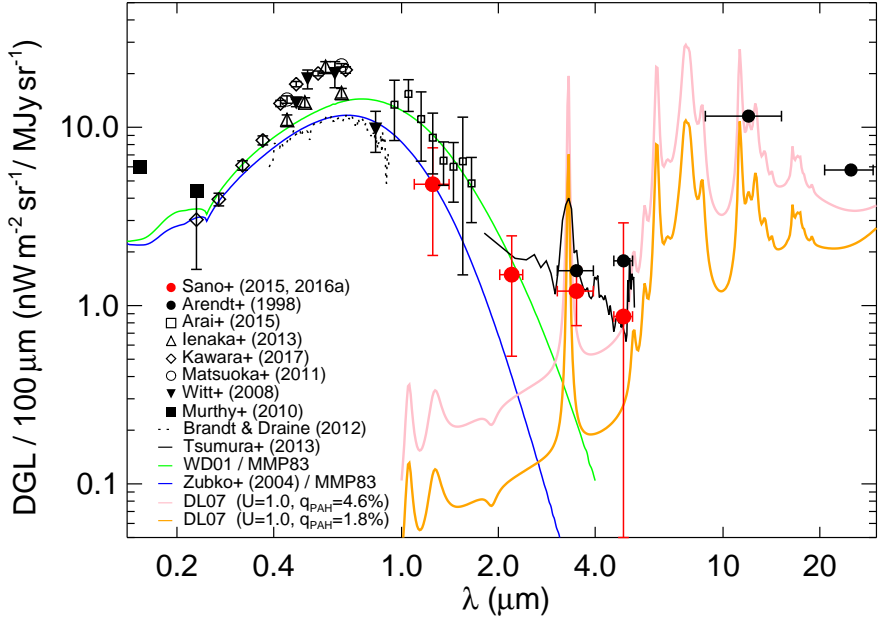} 
  \caption
 {Current results of DGL brightness from UV to mid-IR, scaled by the interstellar $100\,\rm{\mu m}$ emission intensity.
 Observations of wide-field diffuse ISM with {\it COBE}/DIRBE, Cosmic Infrared Background Experiment (CIBER), {\it Hubble Space Telescope} ({\it HST}), {\it Pioneer10/11}, {\it Galaxy Evolution Explorer} ({\it GALEX}), Sloan Digital Sky Survey (SDSS), and {\it AKARI} are indicated by red circles (Sano et al. 2015; 2016a), black filled circles (Arendt et al. 1998), open squares (Arai et al. 2015), open diamonds (Kawara et al. 2017), open circles (Matsuoka et al. 2011), filled squares (Murthy et al. 2010), black dashed curve (BD12), black solid curve (Tsumura et al. 2013), respectively.
 The BD12 result is scaled by a factor of 2.1. 
 Observations toward individual high-$b$ clouds are represented by open triangles (Ienaka et al. 2013) and filled inverse triangles (Witt et al. 2008).
 Model spectra of the scattered light (BD12) are indicated by green and blue curves on the basis of dust models created by WD01 and Zubko et al. (2004), respectively, with the MMP83 ISRF.
 Orange and pink curves represent, respectively, DL07 models of thermal emission assuming $q_{\rm PAH} = 1.8\%$ and $4.6\%$ , with the MMP83 ISRF.
Conversion factor between the intensity ratios ($I_{\lambda, {\rm em}} / I_{100}$) and the DL07 expression ($I_{\lambda, {\rm em}} / N_{\rm H}$) is assumed as $I_{100}/N_{\rm H} = 18.6\pm0.3\,{\rm nW\,m^{-2}\,sr^{-1}}/10^{20}\,{\rm cm^{-2}}$, derived from the DIRBE observation toward $|b| > 25^\circ$  (Arendt et al. 1998). 
For clarity, some symbols are shifted a little from the exact wavelength.
 }
\end{center}
\end{figure*}

Mie (1908) and Debye (1909) formulated the absorption and scattering properties of spherical grains for an electromagnetic wave (Mie theory). 
According to the Mie theory, forward-throwing scattering is dominant in case the wavelength ($\lambda$) is comparable to the grain size ($a$), referred to as ``Mie scattering''.
In contrast, forward and backward scattering become comparable in case of $\lambda \gg a$ (Rayleigh scattering).
A fraction of the scattered intensity toward a scattering angle $\theta$
is represented by a scattering phase function $\Phi_\lambda (\theta)$.
As an indicator of the scattering anisotropy, the first moment of a phase function is defined as
\begin{equation}
g_\lambda \equiv \langle\cos\theta\rangle = \int\Phi_\lambda(\theta)\cos\theta\, d\Omega,
\end{equation}
where $\Omega$ denotes solid angle.
According to this definition, the $g$-factor range is $-1\leq g_\lambda \leq1$ and forward scattering becomes dominant as it is higher.
For interstellar scattering, Henyey \& Greenstein (1941; hereafter HG41) introduced an analytical form of the phase function to be consistent with Equation (1):
\begin{equation}
\phi_\lambda (\theta) = \frac{1}{4\pi} \frac{1-g_\lambda^2}{(1+g_\lambda^2-2g_\lambda \cos\theta)^{3/2}}.
\end{equation}

Considering the scattering anisotropy in the diffuse ISM, Jura (1979; hereafter J79) expected the scattered light intensity as a function of $|b|$, assuming the HG41 phase function and uniform illuminating sources in an infinite Galactic plane.
According to their numerical calculation, intensity ratio of the scattered light ($I_{\lambda,{\rm sca}}$) to interstellar $100\,\rm{\mu m}$ emission ($I_{100}$) is expressed  as
\begin{equation}
\frac{I_{\lambda,{\rm sca}}}{I_{100}} \propto 1-1.1g_\lambda\sqrt{\sin|b|}. 
\end{equation}
This formulation indicates that the $b$-dependence of the intensity ratio becomes steeper as the $g$-factor is higher.

In high-$b$ regions ($|b|\gtrsim20^\circ$), Sano et al. (2016b; hereafter Paper I) found that the intensity ratios of the near-IR DGL ($1.25$ and $2.2\,\rm{\mu m}$) to $100\,\rm{\mu m}$ emission steeply decrease toward the high-$b$ region by analyzing all-sky maps of Diffuse Infrared Background Experiment (DIRBE) on board {\it COBE}.
By assuming the presence of only scattered light, the observed $b$-dependence is fitted by Equation (3) with the $g$-factor of $0.8^{+0.2}_{-0.3}$ at $1.25$ and $2.2\,\rm{\mu m}$.
The derived $g$-factor is too large to explain with the WD01 dust model predicting $ g_\lambda\lesssim 0.3$ in the near-IR: the Rayleigh regime for the WD01 typical grain size ($\sim 0.1\,\rm{\mu m}$).

The steep $b$-dependence can be attributed to the following possibility.
For one thing, the near-IR thermal emission also contributes to the $b$-dependence at $1.25$ and $2.2\,\rm{\mu m}$ due to regional variations of the ISRF and/or the PAH abundance.
Though the near-IR thermal emission has not been found in the diffuse ISM, it has been detected in some reflection nebulae (e.g., Sellgren et al. 1992; Sellgren et al. 1996).
For another, there is room for improving the J79 model of scattered light.
For instance, Draine (2003b; hereafter D03) reported discrepancy of the phase function between the HG41 form and the WD01 dust model in the near-IR.
In addition, J79 suggested a large uncertainty of factor $1.5$ in Equation (3). 
According to these concerns, the high $g$-value derived from the J79 model is unreliable.

In this paper, we adopt models of thermal emission and scattered light to investigate the contribution of these components to the $b$-dependence of the near-IR DGL.
The WD01 model is adopted as interstellar dust properties since it reportedly reproduces the observed  extinction curve (e.g., Fitzpatrick 1999; Draine 2011).
To investigate contribution of the thermal emission, we adopt the DL07 model.
We evaluate contribution of the scattered light, adopting a plane-parallel galaxy model such as BD12. 
The present analysis is also based on the all-sky observation with {\it Planck}.

The remainder of this paper is organized as follows.
Section 2 and 3 describe the models of scattered light and thermal emission, respectively.
In Section 4, we discuss origins of the $b$-dependence of the near-IR DGL by comparing the observation with the models.
Summary and conclusion appear in Section 5.

%% In a manner similar to \objectname authors can provide links to dataset
%% hosted at participating data centers via the \dataset{} command.  The
%% second curly bracket argument is printed in the text while the first
%% parentheses argument serves as the valid data set identifier.  Large
%% lists of data set are best provided in a table (see Table 3 for an example).
%% Valid data set identifiers should be obtained from the data center that
%% is currently hosting the data.
%%
%% Note that AASTeX interprets everything between the curly braces in the 
%% macro as regular text, so any special characters, e.g. "#" or "_," must be 
%% preceded by a backslash. Otherwise, you will get a LaTeX error when you 
%% compile your manuscript.  Special characters do not 
%% need to be escaped in the optional, square-bracket argument.

%% In this section, we use  the \subsection command to set off
%% a subsection.  \footnote is used to insert a footnote to the text.


%% This section also includes several of the displayed math environments
%% mentioned in the Author Guide.

\section{NEAR-INFRARED THEARMAL EMISSION}

Here, we expect intensity ratios of near-IR thermal emission to the $100\,\rm{\mu m}$ emission ($I_{\lambda, {\rm em}} / I_{100}$) as a function of $|b|$, based on {\it Planck} results and the DL07 model.

\subsection{{\it Planck} Observation}

\begin{figure*}
\begin{center}
 \includegraphics[scale=0.5]{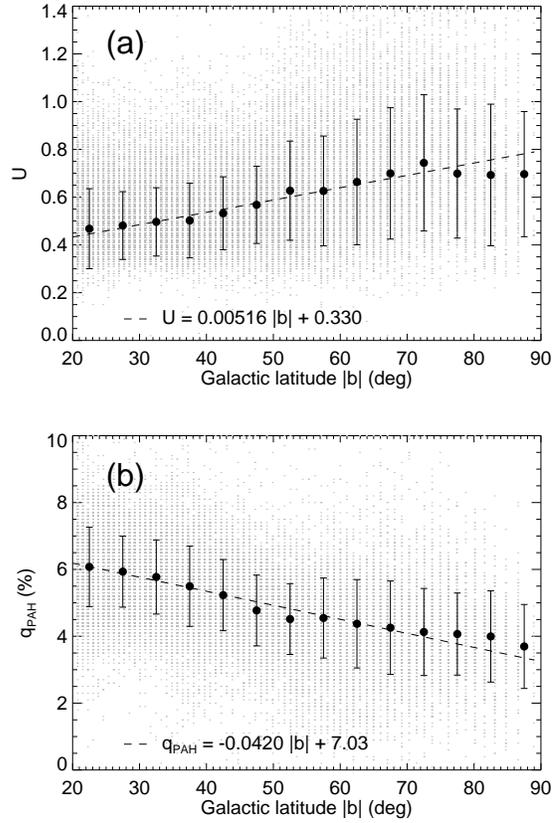} 
  \caption
 {Illustrations of (a) ISRF scaling factor $U (=U_{\rm min})$ and (b) PAH abundance $q_{\rm PAH}$ as a function of $|b|$, taken from the parameter maps of Planck Collaboration et al. (2016).
 In each panel, gray dots represent the parameter values taken from the low-resolution maps of $U_{\rm min}$ and $q_{\rm PAH}$.
Black circles and error bars denote, respectively, the weighted average and standard deviation of data points within certain $|b|$ bins.
 Dashed line indicates a best-fit line to the data points.
 }
\end{center}
\end{figure*}

Planck Collaboration et al. (2016) investigated regional variations of properties of interstellar dust and ISRF from all-sky maps of wide wavelengths range ($10\,\rm{\mu m} \lesssim \lambda \lesssim 1000\,\rm{\mu m}$), including {\it Wide-field Infrared Survey Explorer} ({\it WISE}), {\it IRAS}, and {\it Planck}.
Combining the DL07 model spectra with the all-sky maps, they conducted the spectral energy distribution (SED) fitting in each field of the sky and created maps of interstellar dust and ISRF properties.

In the analysis of thermal emission, crucial parameters in the DL07 model are mass fraction of very small grains with PAH to the total dust, $q_{\rm PAH}(\%)$ and scaling factor of the MMP83 ISRF, $U$.
As $q_{\rm PAH}$ increases, near-IR and mid-IR dust emissions become more dominant. 
In general, the quantity $U$ depends on parameter $\gamma$ indicating a fraction of the photo dissociation region (PDR).
In the DL07 model, a fraction $1-\gamma$ of the dust mass is illuminated by single ISRF intensity $U_{\rm min}$, while the remaining fraction $\gamma$ is heated by various ISRF intensity between $U_{\rm min}$ and $U_{\rm max}$.
As shown in the all-sky maps of each parameter (Figure 1 of Planck Collaboration et al. 2016), the parameter $\gamma$ can be approximated as zero in high-$b$ regions, indicating little contribution of the PDR component.
According to this result, the parameter $U$ can be assumed as $U_{\rm min}$ in high-$b$ regions.
We thus focus on the single ISRF parameter $U=U_{\rm min}$ and $q_{\rm PAH}$ in the DL07 model.

Figure 2 shows $b$-dependence of the parameters $U$ and $q_{\rm PAH}$, taken from the {\it Planck} maps (Planck Collaboration et al. 2016).
The parameter $U$ tends to increase toward the high-$b$ region.
Dust temperature ($T$), another indicator of the ISRF intensity, exhibits the similar trend in the all-sky $T$ map created by Planck Collaboration et al. (2014).
Since the high-$b$ regions are mostly optically thin, interstellar dust may be exposed by the Galactic disk emission intensively.
In contrast, the parameter $q_{\rm PAH}$ tends to decrease toward the high-$b$ region.
Hensley et al. (2016) reported the similar trend from the analysis of the {\it WISE} $12\,\rm{\mu m}$ map.
They showed that ratio of the $12\,\rm{\mu m}$ emission to the dust radiance, an indicator of the PAH abundance, increases toward a low-$b$ region.
In addition, the {\it Planck}-derived value of $q_{\rm PAH} \gtrsim 2\%$ is consistent with the DIRBE observation at $3.5\,\rm{\mu m}$ in high-$b$ regions (Sano et al. 2016a).
Data products created by {\it Planck} are available via Planck Legacy Archive (PLA): https://www.cosmos.esa.int/web/planck/pla.

\subsection{Modeling of the Thermal Emission as a Function of $|b|$}

\begin{figure*}
\begin{center}
 \includegraphics[scale=0.8]{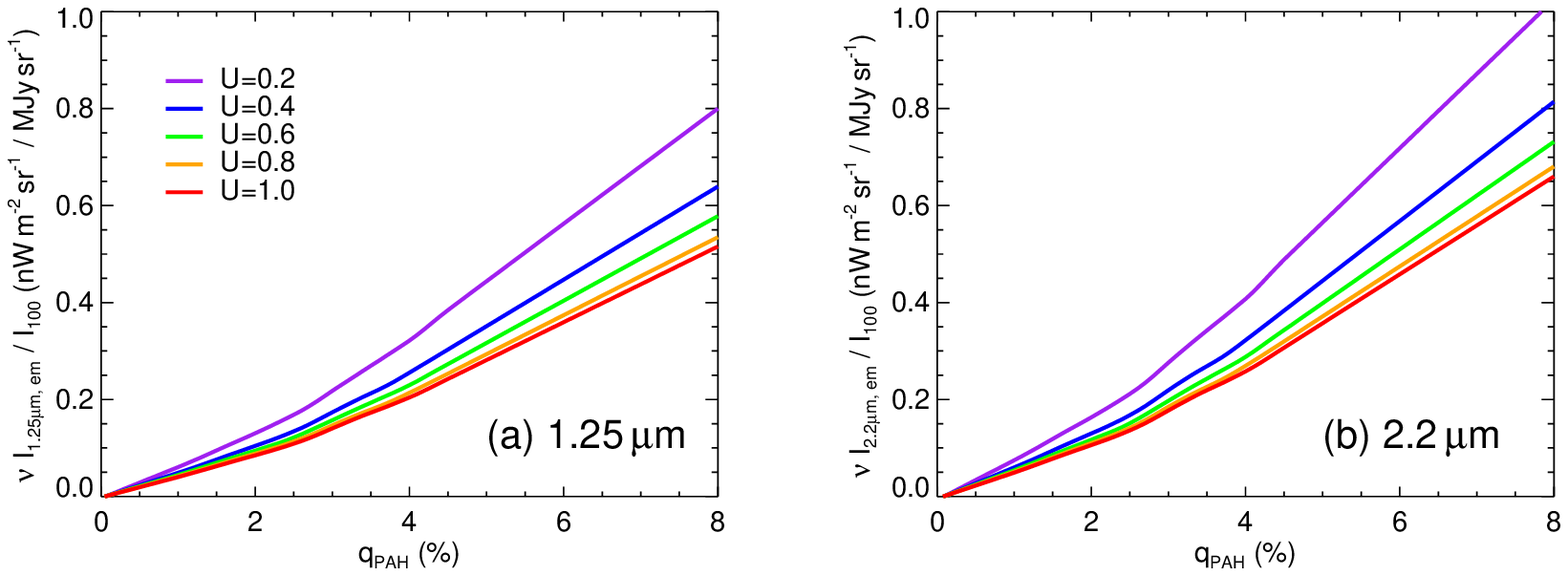} 
 \caption
 {Intensity ratios of near-IR to $100\,\rm{\mu m}$ emission $\nu I_{\lambda, {\rm em}} / I_{100}\,({\rm nW\,m^{-2}\,sr^{-1} / MJy\,sr^{-1}})$ as a function of $q_{\rm PAH}$ in the DL07 model at (a) $1.25\,\rm{\mu m}$ and (b) $2.2\,\rm{\mu m}$.
 In each panel, $\nu I_{\lambda, {\rm em}} / I_{100}$ values at $U=0.2, 0.4, 0.6, 0.8,$ and $1.0$ are represented by purple, blue, green, orange, and red curves, respectively.
}
\end{center}
\end{figure*}

To model the $b$-dependence of $I_{\lambda, {\rm em}} / I_{100}$, we express $U$ and $q_{\rm PAH}$ as a linear function of $|b|$ according to the {\it Planck} observation (Figure 2):
\begin{equation}
U = 0.00516 |b| + 0.330,
\end{equation}
\begin{equation}
q_{\rm PAH} = -0.0420 |b| + 7.03.
\end{equation}
Figure 3 illustrates $\nu I_{\lambda, {\rm em}} / I_{100}$ as a function of $q_{\rm PAH}$ for different values of $U$ in the DL07 model.
The figure is created by interpolating $\nu I_{\lambda, {\rm em}} / I_{100}$ calculated at discrete values of $U$ and $q_{\rm PAH}$ in the DL07 model.

The {\it Planck} observation predicts the parameter values of $0.4 \lesssim U \lesssim 0.8$ and $4\% \lesssim q_{\rm PAH} \lesssim 6\%$ in the high-$b$ region (Figure 2).
In these ranges, the intensity ratios $\nu I_{\lambda, {\rm em}} / I_{100}$ are not sensitive to $U$ but are largely dependent on $q_{\rm PAH}$ (Figure 3).
We then approximate $\nu I_{\lambda, {\rm em}} / I_{100}\,({\rm nW\,m^{-2}\,sr^{-1} / MJy\,sr^{-1}})$ as a function of $q_{\rm PAH}$:
\begin{equation}
\nu I_{1.25\,\rm{\mu m}, {\rm em}}/I_{100} = (0.07\pm0.01)\, q_{\rm PAH},
\end{equation}
\begin{equation}
\nu I_{2.2\,\rm{\mu m}, {\rm em}}/I_{100} = (0.08\pm0.01)\, q_{\rm PAH}.
\end{equation}
By combining the relations (6) and (7) with Equation (5), $\nu I_{\lambda, {\rm em}} / I_{100}$ is expressed as a function of $|b|$.

\section{NEAR-INFRARED SCATTERED LIGHT}

\subsection{Single Scattering in a Plane-Parallel Galaxy}

To estimate $b$-dependence of the scattered light, we employ a plane-parallel galaxy in which the solar system is located in the Galactic plane (BD12).
The model assumes single scattering by dust grains, which is reasonable for the near-IR high-$b$ region.
In the numerical calculation of the scattered light, dust and stellar distributions are expressed as a function of distance $z$ and $z_s$ from the plane, respectively.
Considering dust extinction in a line of sight, the scattered intensity is calculated as
\begin{eqnarray}
I_{\lambda,{\rm sca}} (b) &=& \omega_{\lambda} \csc|b| \int^{\tau_{\lambda}(0)}_0 d\tau_{\lambda} \exp[-\csc|b|(\tau_{\lambda}(0) - \tau_{\lambda})] \nonumber \\ 
&\times&\int^\infty_0 R\, dR \int^{2\pi}_0 d\theta \, \Phi_\lambda \frac{\exp[-A_{\lambda} (z, z_s, R)]}{4\pi[(z-z_s)^2+R^2]} \nonumber \\
&\times&\int^\infty_0 P_{\lambda}(z_s) \,dz_s,
\end{eqnarray}
\begin{equation}
\tau_{\lambda} (z) \equiv \int^\infty_z \sigma_{\rm ext}(\lambda)\rho(z') dz',
\end{equation}
\begin{equation}
A_{\lambda} (z, z_s, R) \equiv |\tau_{\lambda} (z) - \tau_{\lambda} (z_s)|\frac{\sqrt{(z-z_s)^2+R^2}}{|z-z_s|},
\end{equation}
where $\omega_\lambda$ and $\sigma_{\rm ext}(\lambda)$ denote albedo and extinction cross section, respectively.
These dust properties are assumed to be independent of $|b|$.
The quantities $P_{\lambda}(z_s)$ and $\rho(z)$ are surface power density of the stellar sheet and dust density, respectively.
Once these values are supplied, the scattered intensity is calculated by Equation (8) without free parameters.

\subsubsection{Phase Function}

\begin{figure*}
\begin{center}
 \includegraphics[scale=0.8]{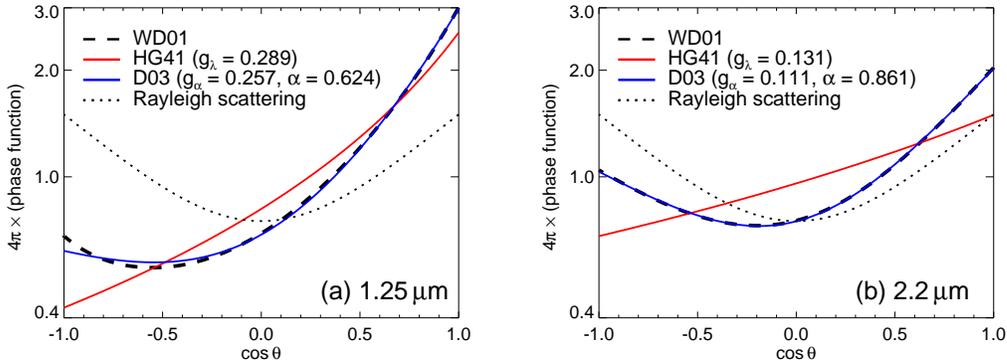} 
 \caption
 {Comparison of the near-IR phase functions as a function of cosine of scattering angle $\theta$ at (a) $1.25\,\rm{\mu m}$ and (b) $2.2\,\rm{\mu m}$.
In each panel, black dashed curve represents the WD01 phase function assuming the $R_V = 3.1$ dust.
Red and blue curves indicate, respectively, the HG41 and D03 phase function formalized by Equation (1) and (11).
The $g$-factor in the HG41 form comes from the first moment of the WD01 phase function.
In the D03 form, the parameters $g_\alpha$ and $\alpha$ are determined to reproduce the WD01 phase function.
A black dotted curve indicates phase function of the Rayleigh scattering.}
\end{center}
\end{figure*}

In the calculation of the scattered light from Equation (8), the phase function $\Phi_\lambda$ should be expressed as a function of scattering angle, such as the HG41 form (Equation 1).
In the near-IR, however, D03 showed that the phase function of HG41 deviates from that of the WD01 dust model assuming the Mie theory.
Therefore, D03 developed a new analytic form to reproduce the WD01 phase function:
\begin{equation}
\phi_{\alpha}(\theta) = \frac{1}{4\pi} \frac{1-g_{\alpha}^2}{(1+g_\alpha^2-2g_\alpha\cos\theta)^{3/2}} \frac{1+\alpha\cos^2\theta}{1+\alpha(1+2g_\alpha^2)/3},
\end{equation}
where $g_\alpha$ and $\alpha$ are adjustable parameters.
In case of $g_\alpha = 0$ and $\alpha = 1$, this form represents the phase function of the Rayleigh scattering.
In case of $\alpha = 0$, it is reduced to the HG41 form (Equation 1).
The parameters $g_\alpha$ and $\alpha$ can be derived by comparing the first and second moments of the WD01 phase function with those of the D03 form.
See Appendices of D03 for detail. 

In Figure 4, the WD01 phase function is compared with the HG41 and D03 forms at $1.25$ and $2.2\,\rm{\mu m}$.
The WD01 phase function shows stronger forward and backward scattering than the HG41 form. 
In contrast, the D03 form well reproduces the WD01 shape in both wavelengths. 
From $1.25$ to $2.2\,\rm{\mu m}$, the WD01 phase function approaches the Rayleigh regime due to the larger difference between the typical grain size and the wavelength.

To compare the modeled $b$-dependence with the DIRBE observation (Paper I), we conduct the above calculation using the WD01 dust properties at $1.22$ and $2.19\,\rm{\mu m}$, close to the two DIRBE bands ($1.25$ and $2.2\,\rm{\mu m}$).
In the adopted WD01 model, the grain abundances are reduced by factor 0.93 from the original one to be consistent with the interstellar extinction (Draine 2003a).
The data of the WD01 models are available at the website: ``www.astro.princeton.edu/$\sim$ draine/''.

\subsubsection{The $100\,\rm{\mu m}$ Emission as a Function of $|b|$}

\begin{figure*}
\begin{center}
 \includegraphics[scale=0.8]{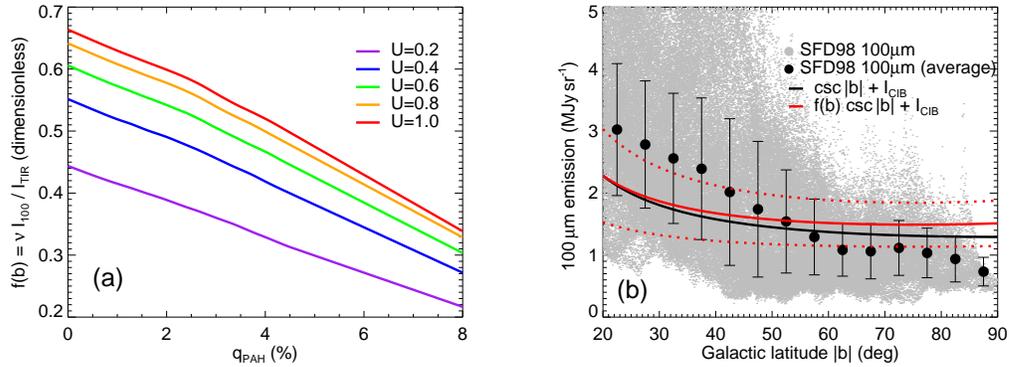} 
 \caption
 {(a): Intensity ratio of $100\,\rm{\mu m}$ emission to the total IR emission $f(b) \equiv \nu I_{100}/I_{\rm TIR}$ as a function of $q_{\rm PAH}$ for different values of $U$ in the DL07 model.
The values $\nu I_{100}/I_{\rm TIR}$ at $U = 0.2, 0.4, 0.6, 0.8,$ and $1.0$ are represented by purple, blue, green, orange, and red curves, respectively.
(b): Comparison of the modeled $I_{100}$ with the observation in high-$b$ regions ($|b|>20^\circ$).
Gray dots represent the $100\,\rm{\mu m}$ intensity expected from the SFD98 map.
Black circles and error bars denote, respectively, the average of the $100\,\rm{\mu m}$ intensity and the standard deviation of data within certain $|b|$ bins in the regions of low $100\,\rm{\mu m}$ intensity ($< 5\,{\rm MJy\,sr^{-1}}$).
Black and red solid curves represent the $I_{100}$ models of $\csc|b| + I_{\rm CIB}$ and $f(b) \csc|b| + I_{\rm CIB}$, respectively, with the isotropic CIB intensity of $I_{\rm CIB} = 0.78\,{\rm MJy\,sr^{-1}}$ (Lagache et al. 2000).
The values of $\csc|b|$ and $f(b) \csc|b|$ are scaled to $1.5\,{\rm MJy\,sr^{-1}}$ at $|b|=20^\circ$.
Red dotted curves indicate the latter model with $f(b)$ of $\pm 50\%$ uncertainty, i.e., $1.5 f(b) \csc|b| + I_{\rm CIB}$ and $0.5 f(b) \csc|b| + I_{\rm CIB}$.
}
\end{center}
\end{figure*}

To predict the $b$-dependence of $I_{\lambda,{\rm sca}}/I_{100}$, we estimate $b$-dependence of the $100\,\rm{\mu m}$ intensity.
In the BD12 model, the $100\,\rm{\mu m}$ intensity can be estimated from the total IR intensity reradiated from dust grains, defined as $I_{\rm TIR} = \int I_{\lambda, {\rm em}} d\lambda$.
In the plane-parallel galaxy model, $b$-dependence of the total IR intensity is expressed as:
\begin{equation}
I_{\rm TIR} \propto \csc|b|.
\end{equation}
To estimate $b$-dependence of $I_{100}$ from $I_{\rm TIR}$, $b$-dependence of $U$ and $q_{\rm PAH}$ should be taken into account because the dust emission SED changes as functions of these quantities in the DL07 model.
The $b$-dependence of $U$ and $q_{\rm PAH}$ can be estimated from the parameter maps of Planck Collaboration et al. (2016).

Now we define a correction factor $f(b)$ as intensity ratio of $100\,\rm{\mu m}$ emission to the total IR emission, which reflects deviation of $I_{100}$ from the simple $\csc|b|$ law:
\begin{equation}
f(b) \equiv \nu I_{100}/I_{\rm TIR}.
\end{equation}
We can estimate $f(b)$ from the dust SED of the DL07 model in combination with the $b$-dependence of $U$ and $q_{\rm PAH}$ (Figure 2).
From Equations (12) and (13), $\nu I_{100}$ is expressed as
\begin{equation}
\nu I_{100} = f(b) I_{\rm TIR} \propto f(b) \csc|b|.
\end{equation}
Therefore, the quantity of interest, $I_{\lambda,{\rm sca}}(b)/I_{100}(b)$ can be written as
\begin{equation}
I_{\lambda,{\rm sca}} (b) / I_{100} (b) = I_{\lambda,{\rm sca}} (b) /  [f(b) I_{\rm TIR} (b)]
\end{equation}
\begin{equation}
\propto I_{\lambda,{\rm sca}} (b) /  [f(b) \csc|b|].
\end{equation}

We estimate the correction factor $f(b)$ from the DL07 model.
Figure 5(a) shows $f(b) \equiv \nu I_{100}/I_{\rm TIR}$ as a function of $q_{\rm PAH}$ for different values of $U$ in the model.
According to the {\it Planck} result (Figure 2), the parameters $U$ and $q_{\rm PAH}$ change, respectively, from $0.4$ to $0.8$ and from $6\%$ to $4\%$ toward high-$b$ regions ($|b|>20^\circ$).
By adapting the $b$-dependence of these parameters to Figure 5(a), the quantity $f(b) \equiv \nu I_{100}/I_{\rm TIR}$ supposedly changes from $0.35$ to $0.5$ when $|b|$ runs from $20^\circ$ to $90^\circ$.
Approximation of this $b$-dependence of $f(b)$ as a linear function of $|b|$ results in
\begin{equation}
f(b) = 0.00214 |b| + 0.307.
\end{equation}
This formula is used in the following analysis.

Figure 5(b) compares the modeled $I_{100}$ with the observation (SFD98).
The $I_{100}$ models are expressed as $\propto \csc|b|$ and $\propto f(b)\csc|b|$.
The SFD98 map should contain the isotropic cosmic infrared background (CIB) component in addition to the interstellar $100\,\rm{\mu m}$ emission.
Therefore, the CIB intensity at $100\,\rm{\mu m}$ of $I_{\rm CIB} = 0.78\pm0.21\,{\rm MJy\,sr^{-1}}$ derived from Lagache et al. (2000) is added to the $I_{100}$ models.
As expected from the functional form of $f(b)$ (Equation 17), the model with the correction factor $f(b)$ shows gentler $b$-dependence than the simple $\csc|b|$ model.

The black circles in Figure 5(b) represent the interstellar $100\,\rm{\mu m}$ intensity averaged in a low-$100\,\rm{\mu m}$ region of less than $5\,{\rm MJy\,sr^{-1}}$.
Since such fields dominate the high-$b$ sky, the averaged intensity is assumed as comparable values for the $I_{100}$ models.
Though the model with the correction factor $f(b)$ should be more preferable from the physical point of view, dispersion of the SFD98 $100\,\rm{\mu m}$ emission is too large to calibrate the $I_{100}$ model.
We thus estimate uncertainty of $f(b)$ in comparison with the SFD98 $100\,\rm{\mu m}$ intensity.
If an ideal model of $I_{100}$ exists, it should be fitted to the averaged values of the SFD98 $100\,\rm{\mu m}$ emission (black dots in Figure 5b).
Therefore, diffrence between the black dots and the model of $f(b)\csc|b| + I_{\rm CIB}$ can be regarded as uncertainty of the correction factor $f(b)$.
In Figure 5(b), red dotted curves represent the $I_{100}$ model with $\pm50\%$ uncertainty of $f(b)$, i.e., $1.5f(b)\csc|b| + I_{\rm CIB}$ and $0.5f(b)\csc|b| + I_{\rm CIB}$.
The values of black dots are approximately within these two curves throughout the high-$b$ region.
We thus evaluate the $f(b)$ uncertainty as $\sim \pm 50\%$.

\subsubsection{Vertical Distributions of Interstellar Dust and Stars}

In the calculation of the scattered light, we assume functional forms of $\rho(z)$ and $P_{\lambda}(z_s)$ according to observations of vertical distributions of interstellar dust and stars.
Numerous studies have investigated the interstellar dust distribution (e.g., Lyng\aa 1982; M\'endez \& van Altena 1998; Malhotra 1995; Nakanishi \& Sofue 2003).
Taking into account these observations, we adopt the following two forms of $\rho(z)$.
One is a Gaussian distribution with its variance of $\sigma_1 = 250\,{\rm pc}$ (Malhotra 1995; Nakanishi \& Sofue 2003) and another is an exponentially-decreasing density with its scale height of $\sigma_2 = 110\,{\rm pc}$ (Lyng\aa 1982).
Therefore, they are expressed as
\begin{equation}
\rho_1(z) \propto \exp(-z^2/2\sigma_1^2),
\end{equation}
\begin{equation}
\rho_2(z) \propto \exp(-z/\sigma_2).
\end{equation}
The dust density following the formula (19) decreases more steeply toward the vertical direction than that following the formula (18).

Similarly, a number of studies have explored the vertical distribution of Galactic stars (e.g., Rana \& Basu 1992; Binney \& Merrifield 1998; Gilmore \& Reid 1983; Girardi et al. 2005).
In the present analysis, we assume the following three $P_{\lambda}(z_s)$, namely, Case 1, Case 2, and Case 3.
In Case 1, stellar distribution is expressed as a sum of two exponential functions:
\begin{equation}
P_{1,\lambda}(z_s) \propto 0.9\exp(-z_s/h_1) + 0.1\exp(-z_s/h_2),
\end{equation}
with $h_1 = 300\,{\rm pc}$ and $h_2 = 1350\,{\rm pc}$, corresponding to the scale height of the thin and thick disk, respectively (Binney \& Merrifield 1998; Gilmore \& Reid 1983).
This model is the same as that adopted in the previous calculation of BD12.

In Case 2, the stellar distribution is assumed as a linear combination of the squared hyperbolic secant:
\begin{equation}
P_{2,\lambda}(z_s) \propto 0.9 \, {\rm sech}^2 [0.5z_s/h(t)] + 0.1 \, {\rm sech}^2 (0.5z_s/h).
\end{equation}
This form is preferred by Girardi et al. (2005), who developed a star-counts model assuming a three-dimensional stellar distribution, age-metallicity relation, and star-formation rate in the Milky Way.
In the formula (21), scale height of the thin disk $h(t)$ is expressed as a function of the stellar age $t$, based on observational study of the age-metallicity relation of various stars (Rana \& Basu 1992):
\begin{equation}
h(t) = z_0 (1+t/t_0)^\beta, 
\end{equation}
where $z_0$, $t_0$, and $\beta$ are adjustable parameters.
Girardi et al. (2005) calibrated the model according to the real star-counts and determined the parameters of $z_0 = 94.7\,{\rm pc}$, $t_0 = 5.55\,{\rm Gyr}$, and $\beta = 1.67$, which are adopted in the present analysis.
To investigate sensitivity of the $b$-dependence of the scattered light to the stellar age, we set $t$ as $1$, $5$, and $10\,{\rm Gyr}$ in each calculation, which correspond to $h(t) = 125$, $277$, and $529\,{\rm pc}$, respectively (Equation 22).
In the near-IR, red stellar population classified as $K$- or $M$-type stars of $t \gtrsim 10\,{\rm Gyr}$ are thought to dominate the sky brightness of the Milky Way.
Scale height of the thick disk is set to $h = 800\,{\rm pc}$ since it supposedly comprises the old stellar population of $t \gtrsim 10\,{\rm Gyr}$ (Girardi et al. 2005). 
The density fraction of the thick disk to the thin disk is assumed as $10\%$, same as the formula (20).
This fraction is marginally consistent with the star-counts model (Girardi et al. 2005), indicating dominant contribution of the thin disk to the $b$-dependence of the scattered light.

In Case 3, all stars are assumed to exist in the Galactic plane, which corresponds to $z_s = 0$ in Equation (8).
The extreme and unrealistic stellar distribution was adopted in the previous estimation of the $b$-dependence of the scattered light (J79).

\subsection{Scattered Light in a Dusty Slab}

In addition to the single scattering, we evaluate effect of multiple scattering.
Multiple scattering is usually treated by a Monte Carlo simulation according to the theory of random numbers.
Using the simulation, several studies have investigated the multiple scattering in diffuse ISM or clumpy media, such as reflection nebulae (e.g., Witt 1977; Witt \& Gordon 1996; Murthy 2016).
Fortunately in the Galactic scale, an analytic form of the scattered light intensity is present as a solution of radiative transfer in a dusty slab (HG41).
Since such an analytic solution is easy to deal with, we adopt the scattered light model to estimate the effect of the multiple scattering.

According to HG41, differential equations of radiative transfer of starlight $I_{\lambda,{\rm star}}$ and scattered light $I_{\lambda,{\rm sca}}$ through a dusty slab are expressed as
\begin{equation}
\cos\Theta \frac{dI_{\lambda,{\rm star}}}{d\tau_\lambda} = I_{\lambda,{\rm star}} - a_\lambda,
\end{equation}
\begin{equation}
\cos\Theta \frac{dI_{\lambda,{\rm total}}}{d\tau_\lambda} = I_{\lambda,{\rm total}} - \int \omega_\lambda I_{\lambda,{\rm total}} \, \Phi_\lambda \, d\Omega - a_\lambda,
\end{equation}
\begin{equation}
I_{\lambda,{\rm total}} = I_{\lambda,{\rm star}} + I_{\lambda,{\rm sca}},
\end{equation}
where $\Theta$ is an angle from the vertical direction of the Galaxy (i.e., $\Theta = 90^\circ - |b|$).
The quantity $a_\lambda$ is ratio of stellar emission to absorption coefficient.
By adopting the HG41 phase function (Equation 1) and several approximations, the scattered light intensity is expressed as
$$
I_{\lambda,{\rm sca}} = \frac{a_\lambda}{1-\omega_\lambda}\biggl[\omega_\lambda + (1-\omega_\lambda) \exp(-\tau_{\lambda,\infty} \sec\Theta)
$$
$$
- \exp[-\tau_{\lambda,\infty} (1-\omega_\lambda g_\lambda) \sec\Theta]
- \frac{\omega_\lambda (1-g_\lambda)}{1-\omega_\lambda g_\lambda} 
$$
$$
\times \frac{1-\exp[-\tau_{\lambda,\infty} (1-\omega_\lambda g_\lambda) \sec\Theta]}{1-\frac{3(1-\omega_\lambda)}{1-\omega_\lambda g_\lambda} \cos^2 \Theta} 
$$
\begin{equation}
\times \frac{\cosh p_\lambda \tau_{\lambda,\infty} + \frac{p_\lambda}{1-\omega_\lambda g_\lambda} \cos\Theta \sinh p_\lambda \tau_{\lambda,\infty}}{\cosh p_\lambda \tau_{\lambda,\infty} + \frac{2p_\lambda}{3(1-\omega_\lambda g_\lambda)} \sinh p_\lambda  \tau_{\lambda,\infty}}\biggr],
\end{equation}
\begin{equation}
p_\lambda^2 = 3(1-\omega_\lambda)(1-\omega_\lambda g_\lambda),
\end{equation}
where $\tau_{\lambda,\infty}$ is defined as half-thickness of the slab in a plane-parallel galaxy, i.e., optical thickness toward $|b| = 90^\circ$ direction.
Probability of the multiple scattering is thought to be higher as the quantity $\tau_{\lambda,\infty}$ increases.
Therefore, effect of the multiple scattering can be assessed by changing values of $\tau_{\lambda,\infty}$.

\section{RESULT AND DISCUSSION}

\subsection{Contribution of the Thermal Emission Component}

\begin{figure*}
\begin{center}
 \includegraphics[scale=0.8]{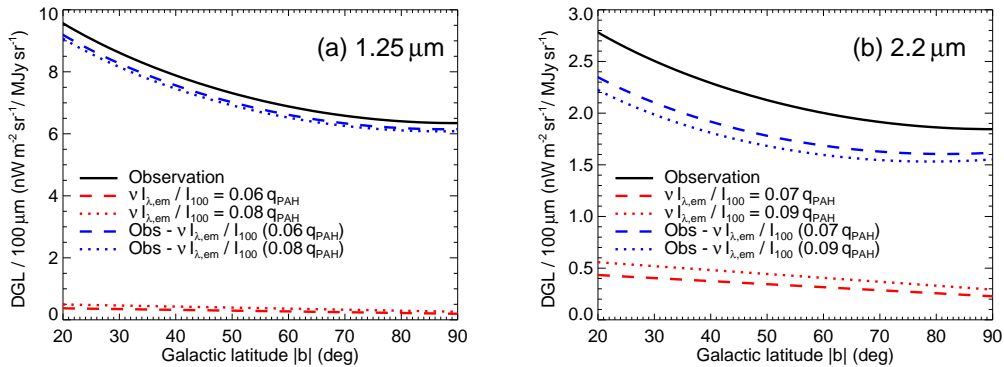} 
 \caption
 {Comparison of the near-IR thermal emission and the observed $b$-dependence of the DGL at (a) $1.25\,\rm{\mu m}$ and (b) $2.2\,\rm{\mu m}$.
 The observed $b$-dependence is indicated by black solid curve following the J79 form with $1$-sigma lower limit of $g_\lambda=0.5$ (Paper I).
 Red dashed and dotted lines represent, respectively, the $b$-dependence of $\nu I_{\lambda,{\rm em}}/I_{100}$ assuming the minimum and maximum cases of the relation between $\nu I_{\lambda,{\rm em}}/I_{100}$ and $q_{\rm PAH}$ (Equations 6 and 7).
Corresponding residual components (i.e., scattered light) are indicated by blue dashed and dotted curves.}
\end{center}
\end{figure*}

Using the relation between $\nu I_{\lambda,{\rm em}}/I_{100}$ and $|b|$ as described in Section 2, we compare the thermal emission component with the observed $b$-dependence (Paper I).
Figure 6 shows relative contribution of the thermal emission component.
The red dashed and dotted lines represent, respectively, the minimum and maximum cases in Equation (6) and (7), i.e., $\nu I_{\lambda,{\rm em}}/I_{100} = 0.06\,q_{\rm PAH}\,(0.07\,q_{\rm PAH})$ and $0.08\,q_{\rm PAH}\,(0.09\,q_{\rm PAH})$ at $1.25\,\rm{\mu m}$ ($2.2\,\rm{\mu m}$).
Also, the residual scattered light component derived by subtracting the thermal emission from the observed DGL is indicated by the blue dashed and dotted curves.

In both bands, difference between the minimum and maximum cases is small enough to regard them as identical, considering uncertainties associated with the observation.
At $1.25\,\rm{\mu m}$, the thermal emission component is less than $10\%$ of the total DGL, which is also implied from Figure 1.
Therefore, the observed $b$-dependence can be largely attributed to the scattered light component in this band.
At $2.2\,\rm{\mu m}$, contribution of the thermal emission is higher than that at $1.25\,\rm{\mu m}$ and the ratio to the total DGL is $\sim 20\%$.
As shown in Figure 6, a little contribution of $\nu I_{\lambda,{\rm em}}/I_{100}$ makes the $b$-dependence of the residual scattered light rather gentler in the high-$b$ region.
However, the near-IR thermal emission component expected from the DL07 model does not fully account for the observation.

\subsection{Contribution of the Scattered Light Component}

On the basis of the $b$-dependence of the scattered light (Figure 6), we search for the origin in comparison with the various scattered light models described in Section 3.
To focus on steepness of the $b$-dependence, results of $\nu I_{\lambda,{\rm sca}}/I_{100}$ are scaled to unity at $|b|=20^\circ$.

\subsubsection{Effect of the Different Phase Function}

To investigate contribution of the different forms of phase function (HG41 or D03) to the $b$-dependence, the $100\,\rm{\mu m}$ intensity is assumed to be proportional to $\csc|b|$ without the correction factor $f(b)$.
In the single scattering model (Equation 8), dust density is set as the Gaussian form.
The stellar distribution is assumed as Case 1 or Case 3.

Figure 7 compares the $b$-dependence of $\nu I_{\lambda,{\rm sca}}/I_{100}$ expected from the HG41 and D03 phase function.
At both $1.25$ and $2.2\,\rm{\mu m}$, the model assuming the D03 phase function is closer to the observed $b$-dependence by more than $10\%$.
This trend is rather distinctive at $2.2\,\rm{\mu m}$.
This may indicate that both forward and backward scattering are critical to make steeper $b$-dependence of the scattered light as seen in the shapes of the phase function (Figure 4).
However, this scattered light model is not enough to explain the observed steepness.

Among the scattered light models illustrated in Figure 7, the model assuming the D03 phase function and Case 1 should be most close to the real situation in terms of the current understanding of interstellar dust properties and vertical structures of the Milky Way.
We thus assume the model as default one in the following discussion.

In Figure 7, the models adopting Case 3 and the HG41 phase function (red dashed curves) should be close to the J79 form (Equation 3).
In comparison, the J79 form is represented by black dotted curves with the $g$-factor set to the first moment of the WD01 phase function, i.e., $0.289$ and $0.131$ at $1.22$ and $2.19\,\rm{\mu m}$, respectively.
The difference between our calculation and the J79 model is within $\sim 10\%$ in both bands.
The discrepancy may be caused by different assumption of the dust distribution: J79 did not consider the density gradient toward the vertical direction since they assumed a single high-$b$ cloud.

\begin{figure*}
\begin{center}
 \includegraphics[scale=0.8]{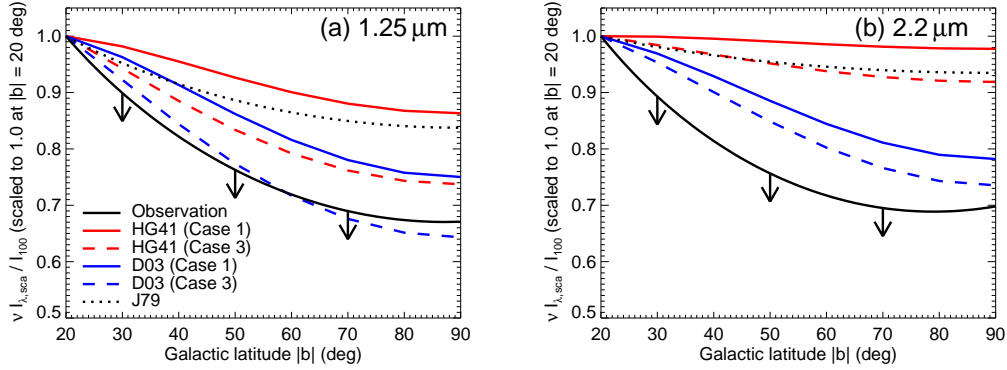} 
 \caption
 {Comparison of the observed and modeled $\nu I_{\lambda,{\rm sca}}/I_{100}$ at (a) $1.25\,\rm{\mu m}$ and (b) $2.2\,\rm{\mu m}$.
In each panel, black solid curve indicates the $b$-dependence derived by subtracting the thermal emission component $\nu I_{\lambda,{\rm em}}/I_{100}$ from the observed $b$-dependence of the DGL (blue dotted curves in Figure 6).
The solid (dashed) red and blue curves represent, respectively, the Case 1 (3) results assuming the HG41 and D03 phase function in the single scattering model (see text).
A black dotted curve indicates the previous model (J79; Equation 3).
All are scaled to unity at $|b|=20^\circ$.}
\end{center}
\end{figure*}

\subsubsection{Effect of the $b$-dependence of the $100\,\rm{\mu m}$ Emission}

Figure 8 compares the quantities $\nu I_{\lambda,{\rm sca}}/I_{100}$ assuming the default model of the scattered light with or without the $I_{100}$ correction factor $f(b)$.
At both $1.25$ and $2.2\,\rm{\mu m}$, models with $f(b)$ exhibit steeper $b$-dependence.
This is because the $b$-dependence of $I_{100}$ with $f(b)$ are gentler than that without $f(b)$ by a factor of $1.5$ at the maximum (Figure 5b).
As a result, the models of $I_{\lambda,{\rm sca}}/I_{100}$ can account for the observed $b$-dependence in most of the high-$b$ region.
At this stage, the observed $b$-dependence of $I_{\lambda,{\rm sca}}/I_{100}$ can be reproduced by the scattered light models assuming the recent interstellar dust properties (WD01) and the $\csc|b|$-corrected $I_{100}$ based on the {\it Planck} observation.

This result indicates that the quantity $I_{\lambda,{\rm sca}}/I_{100}$ is sensitive to the $b$-dependence of $I_{100}$.
As estimated in Section 3.1.2, the correction factor $f(b)$ includes $\sim \pm50\%$ uncertainty in comparison with the $b$-dependence of the SFD98 map of the $100\,\rm{\mu m}$ emission.
Due to the large uncertainty of the $100\,\rm{\mu m}$ emission in terms of the $b$-dependence, it may be useful to investigate other normalizing quantities of the DGL, e.g., optical depth at $100\,\rm{\mu m}$.
Such study will be helpful for more robust analysis of the DGL.

\begin{figure*}
\begin{center}
 \includegraphics[scale=0.8]{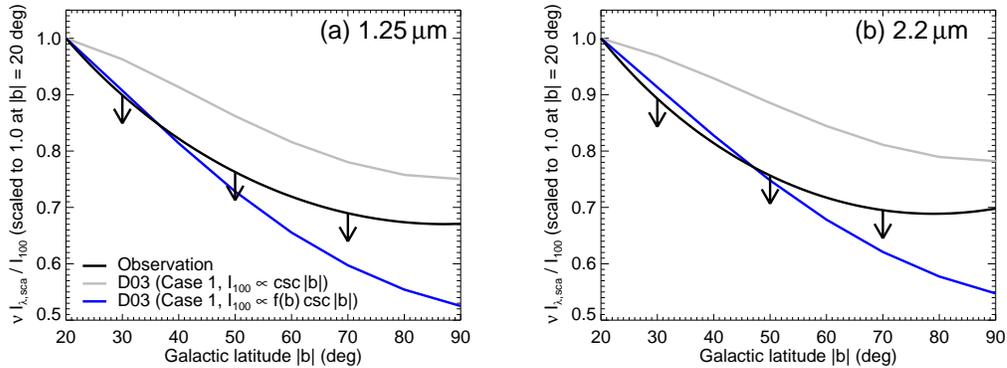} 
 \caption
 {Same as Figure 7, but comparison of the $b$-dependence of $\nu I_{\lambda,{\rm sca}}/I_{100}$ with or without the $I_{100}$ correction factor $f(b)$.
 The scattered light models assuming the D03 phase function and Case 1 scaled by $I_{100} \propto \csc|b|$ are indicated by gray curves (i.e., the blue solid curves in Figure 7), while those scaled by $I_{100} \propto f(b)\csc|b|$ are represented by blue solid curves.
 }
\end{center}
\end{figure*}

\subsubsection{Effect of the Vertical Structure of the Milky Way}

To estimate sensitivity of the $b$-dependence to the vertical distributions of interstellar dust and stars, we compare the single scattering models assuming the various functional forms (Section 3.1.3).
Figure 9 compares the various scattered light models calculated by the two dust density with the three stellar distributions of the different stellar age.
These results are scaled by the default model of the scattered light.

In terms of the effect on the $b$-dependence, the stellar age (scale height) of the thin disk is more influential than the dust distribution (Figure 9).
As the scale height increases, the relative ratio rises toward the high-$b$ region, indicating the $b$-dependence of the scattered light is gentler.
Deviations of the individual models from the default one is within $\sim 5\%$ throughout the high-$b$ region.
Considering the real situation of mixture of various stellar population, the deviation may be smaller than $5\%$.
This result indicates that the variety in the assumed vertical distribution of the Milky Way is less influential to the $b$-dependence than the other factors discussed in Section 4.2.1 and 4.2.2.

\begin{figure*}
\begin{center}
 \includegraphics[scale=0.8]{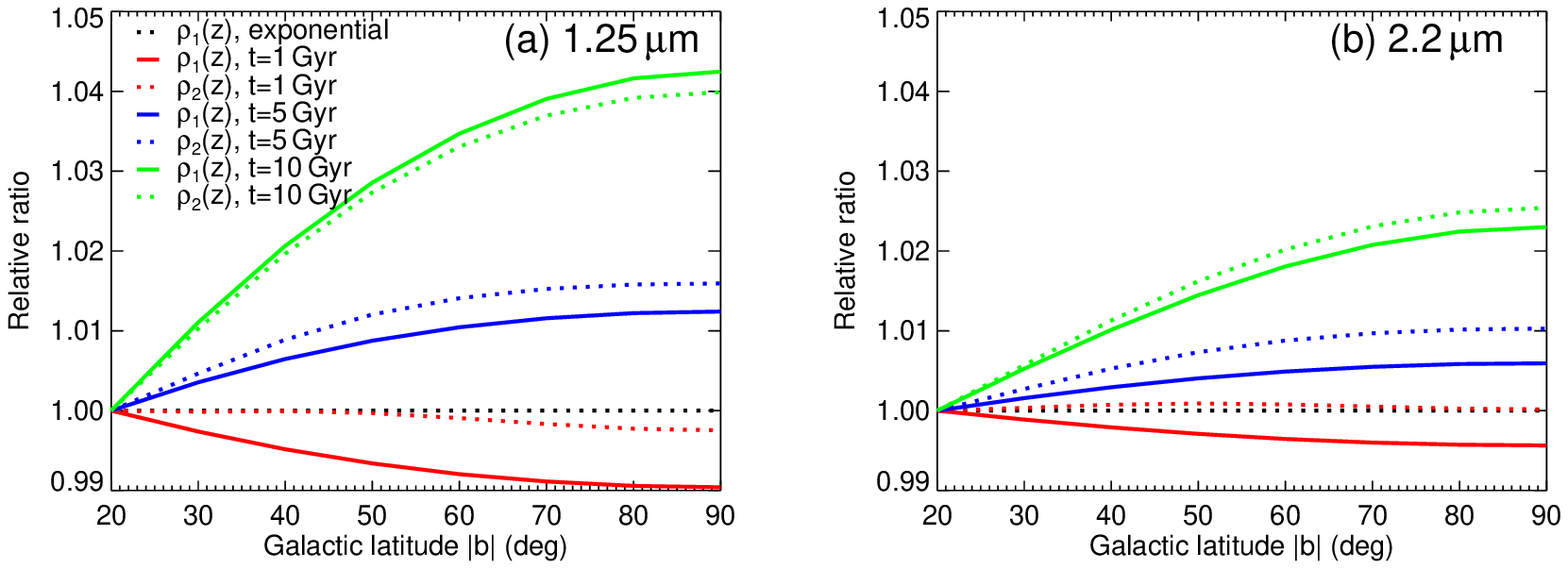} 
 \caption
 {Comparison of the single scattering models as a function of $|b|$, assuming different vertical distributions of interstellar dust and stars at (a) $1.25\,\rm{\mu m}$ and (b) $2.2\,\rm{\mu m}$.
 Each result is normalized by the default models assuming $\rho_1(z)$, $P_{1,\lambda}(z_s)$, and the D03 phase function (black dotted line).
Individual curves represent the models assuming $P_{2,\lambda}(z_s)$ of the stellar age $t = 1$ (red), $5$ (blue), and $10\,\rm{Gyr}$ (green) with the dust density of $\rho_1(z)$ (solid) and $\rho_2(z)$ (dotted).
}
\end{center}
\end{figure*}

\subsubsection{Effect of the Multiple Scattering}

Effect of the multiple scattering is estimated by using the analytic form of the scattered light intensity (Section 3.2).
According to the definition of the half-thickness $\tau_{\lambda,\infty}$, large $\tau_{\lambda,\infty}$ is expected to cause the high probability of the multiple scattering.
Figure 10 illustrates the near-IR $b$-dependence of $\nu I_{\lambda,{\rm sca}}/I_{100}$ derived from Equation (26) and $I_{100} \propto f(b) \csc|b|$, along with that in the near-UV ($0.3\,\rm{\mu m}$) and optical $V$ band ($0.55\,\rm{\mu m}$).
In Equation (26), the albedo and $g$-factor at each wavelength are taken from the WD01 dust model assuming $R_V = 3.1$.
In high-$b$ regions, the visual optical depth expected from the SFD98 reddening map is approximately within $\tau_V = 0.05\csc|b|$ and $0.15\csc|b|$ (Figure 11 of BD12).
To be consistent with the reddening map, we set the quantity $\tau_{V,\infty}$ as $0.05$, $0.10$, and $0.15$ in each wavelength.
To derive the corresponding half-thickness $\tau_{\lambda,\infty}$ in the other wavelengths, extinction curve expected from the WD01 model is used.

As shown in Figure 10, $b$-dependence of $\nu I_{\lambda,{\rm sca}}/I_{100}$ assuming the different $\tau_{\lambda,\infty}$ becomes similar toward longer wavelengths, due to the lower possibility of the multiple scattering.
At $1.25$ and $2.2\,\rm{\mu m}$, difference among the models assuming the three $\tau_{\lambda,\infty}$ is within a few percent throughout the high-$b$ region.
Therefore, it is likely that the $b$-dependence of the near-IR scattered light is nearly independent on the multiple scattering.

The analytic form of the scattered light is based on the HG41 phase function.
In comparison with the analytic model, the single scattering model adopting the HG41 phase function is plotted in Figure 10(c) and (d).
Difference between the models of single and multiple scattering is approximately less than a few percent.
We thus conclude that the effect of the multiple scattering probably does not contribute to the $b$-dependence of the near-IR scattered light.

\begin{figure*}
\begin{center}
 \includegraphics[scale=0.8]{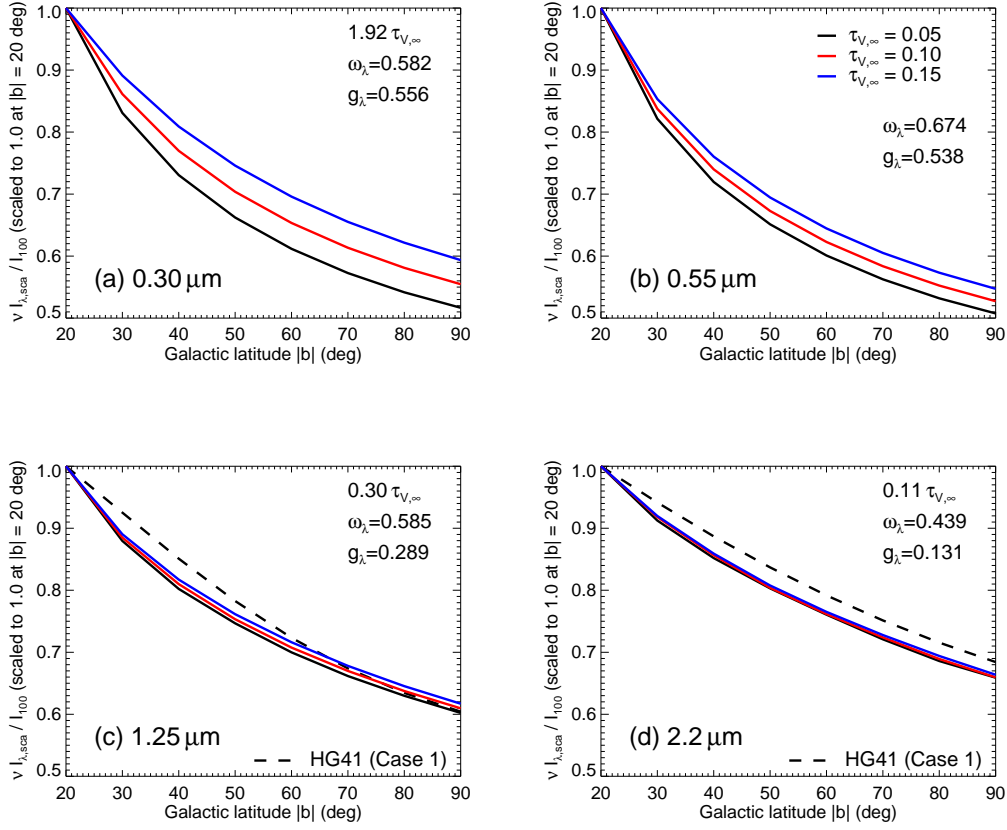} 
 \caption
 {Comparison of $\nu I_{\lambda,{\rm sca}}/I_{100}$ as a function of $|b|$, expected from the analytic form of the scattered light intensity with $I_{100} \propto f(b) \csc|b|$ in the four wavelengths; (a) $0.30$, (b) $0.55$, (c) $1.25$, and (d) $2.2\,\rm{\mu m}$.
In each panel, black, red, and blue solid curves represent, respectively, the analytic models with the $0.55\,\rm{\mu m}$ half-thickness of $\tau_{V,\infty} = 0.05$, $0.10$, and $0.15$.
In the other wavelengths, $\tau_{\lambda,\infty}$ is estimated from $\tau_{V,\infty}$ by using the interstellar extinction of the WD01 dust model with the scaling factor relative to $\tau_{V,\infty}$ shown in the upper right part of each panel.
Black dashed curves in the panels (c) and (d) indicate the result of the single scattering models assuming $\rho_1(z)$, $P_{1,\lambda}(z_s)$, and the HG41 phase function.
}
\end{center}
\end{figure*}

\subsection{Possible Contributions of Other Factors}

\subsubsection{Fluctuation of the Cosmic Infrared Background}

Particularly in high-$b$ regions, the SFD98 $100\,\rm{\mu m}$ map is reportedly influenced by the CIB component which is not associated with the interstellar dust emission (Yahata et al. 2007; Meisner \& Finkbeiner 2015).
Several studies have claimed that the CIB observed in both near-IR and far-IR shows spatial fluctuation due to the galaxy clustering or other hypothetical sources including first stars, intra halo light associated with outer galaxies, or direct collapse black holes in the early universe (e.g., Lagache et al. 2007; Matsumoto et al. 2011; Matsuura et al. 2011; Cooray et al. 2012; Yue et al. 2013; Zemcov et al. 2014).
If the CIB fluctuation in the near-IR and far-IR correlates with each other, it possibly affects the intensity ratio of the near-IR DGL to the interstellar $100\,\rm{\mu m}$ emission.

Typical angular scale of the fluctuation created by the extragalactic sources are expected to be less than an order of $1^\circ$, while that of radiation from interstellar dust (i.e., DGL or far-IR emission) is thought to be larger than that scale (e.g., Lagache et al. 2007; Matsuura et al. 2011).
Therefore, the effect of the CIB fluctuation should be taken into account if the DGL analysis is conducted in the angular scale of $\lesssim 1^\circ$.
However, we focus on much larger scale of $\gtrsim 10^\circ$ in the correlation analysis of the DGL and $100\,\rm{\mu m}$ emission (e.g, Matsuoka et al. 2011; BD12; Paper I).
In the large scale, contribution of the CIB fluctuation is presumably small enough to regard the CIB component as uniform.
Therefore, it is unlikely that the CIB fluctuation is influential in the analysis of the $b$-dependence of the DGL.

\subsubsection{Size and Shape of Dust Grains}

Though the present analysis is based on the WD01 dust model, several studies have suggested the presence of $\rm{\mu m}$-sized large grains in addition to the WD01 dust.
For example, Wang et al. (2015) added the $\rm{\mu m}$-sized grains to the WD01 model and reproduced the flat extinction curve observed in $\sim 3$--$10\,\rm{\mu m}$  (Wang et al. 2013; Nishiyama et al. 2009; Gao et al. 2009; Flaherty et al. 2007; Jiang et al. 2006; Indebetouw et al. 2015; Lutz 1999).
Notably, this modification does not violate the observed extinction curve from UV to near-IR.
The presence of large grains has also been suggested by the derivation of the high albedo in the near-IR (Block et al. 1994; Witt et al. 1994; Lehtinen \& Mattila 1996).
The large grain population is expected to cause a stronger forward-throwing phase function in the near-IR since the Mie scattering becomes more dominant.
This effect may also influence the $b$-dependence of the scattered light.

In addition to the controversy in the dust size, there is no guarantee that the interstellar dust grains are spherical.
In the calculation of scattering anisotropy, the Mie theory cannot be applied to the nonspherical dust. 
To estimate the scattering properties of such grains including porous dust aggregates, several studies have developed various numerical methods, such as the discrete dipole approximation and the $T$-matrix method (e.g., Purcell \& Pennypacker 1973; Draine \& Flatau 1994; Mishchenko et al. 1996; Tazaki et al. 2016).
These effects on the $b$-dependence will be investigated in the future.

\section{SUMMARY AND CONCLUSION}

To reveal the origin of the steep $b$-dependence of the intensity ratios of near-IR ($1.25$ and $2.2\,\rm{\mu m}$) DGL to interstellar $100\,\rm{\mu m}$ emission, we present the analysis according to the models of thermal emission and scattered light with the assistance of the {\it Planck} observation. 

We predict the intensity ratios of the interstellar near-IR to $100\,\rm{\mu m}$ emission as a function of $|b|$, using the DL07 dust emission model and the $b$-dependence of the PAH abundance derived from {\it Planck}.
We find the intensity ratio increasing toward the low-$b$ region, but the contribution of the thermal emission to the observed DGL is less than $\sim20\%$ at both $1.25$ and $2.2\,\rm{\mu m}$.
We then proceed the analysis in terms of the $b$-dependence of the scattered light component.

To express the scattered light intensity as a function of $|b|$, we adopt a plane-parallel galaxy model in which single scattering occurs according to vertical structures of interstellar dust and stars.
Since the classical HG41 phase function reportedly deviates from the recently developed dust model (WD01), we modify the form according to the D03 approximation.
We also evaluate the $b$-dependence of the $100\,\rm{\mu m}$ emission by applying the correction factor to the simple $\csc|b|$ law, based on the regional variations of the PAH abundance and the ISRF intensity derived from the {\it Planck} observation.
We find that the models assuming these factors cause the steeper $b$-dependence of the intensity ratio of the scattered light to the $100\,\rm{\mu m}$ emission and it can account for the observed steep $b$-dependence.
However, the correction factor of the $100\,\rm{\mu m}$ emission includes large uncertainty of $\sim \pm50\%$ in comparison with the observed dispersion of the $100\,\rm{\mu m}$ emission.
As future work, it will be useful to find more robust tracer of the DGL, if it exists.

In addition to these two factors, we investigate effects of various assumptions of the vertical structures and multiple scattering by taking into account these factors in the calculation of the scattered light.
As a result, these two factors are less influential to the $b$-dependence than the corrections of the phase function and $100\,\rm{\mu m}$ emission.

In conclusion, the observed $b$-dependence of the near-IR DGL can be explained by the scattering anisotropy expected from the recent interstellar dust model with a little contribution of the near-IR thermal emission based on the same dust model.
Present analysis thus suggests that the recent interstellar dust model (WD01) is successful in accounting for the $b$-dependence of the near-IR DGL as well as other observations of interstellar dust.

\acknowledgments
We thank the referee for a number of constructive comments that improved the manuscript.
This work is based on observations obtained with {\it Planck} (http://www.esa.int/Planck), an ESA science mission with instruments and contributions directly funded by ESA Member States, NASA, and Canada.
K.S. was a Research Fellow of Japan Society for the Promotion of Science (JSPS) Grant Number 02607628.
S.M. was supported by JSPS KAKENHI Grant Number 15H05744.


\begin{thebibliography}{}
\bibitem[Arai(2015)]{arai2015} Arai, T., Matsuura, S., Bock, J., et al.\ 2015, \apj, 806, 69
\bibitem[Arendt(1998)]{arendt1998} Arendt, R.~G., Odegard, N., Weiland, J.~L., et al.\ 1998, \apj, 508, 74
\bibitem[Binney(1998)]{binney1998} Binney, J., \& Merrifield, M. \ 1998, Galactic Astronomy (Princeton, NJ: Princeton Univ. Press)
\bibitem[Block(1994)]{block1994} Block, D.~L., Witt, A.~N., Grosbol, P., et al.\ 1994, \aap, 288, 383
\bibitem[Brandt \& Draine(2012)]{brandt2012} Brandt, T.~D., \& Draine, B.~T.\ 2012, \apj, 744, 129
\bibitem[Cooray(2012b)]{cooray2012b} Cooray, A., Smidt, J., de Bernardis, F., et al. \ 2012b, \nat, 490, 514
\bibitem[Debye(1909)]{debye1909} Debye, P. \ 1909, Anp, 335, 57 
\bibitem[Draine(2011)]{draine20011} Draine, B.~T.\ 2011, Physics of the Interstellar and Intergalactic Medium (Princeton, NJ: Princeton Univ. Press)
\bibitem[Draine(2003a)]{draine2003a} Draine, B.~T.\ 2003a, ARA\&A, 41, 241
\bibitem[Draine(2003b)]{draine2003b} Draine, B.~T.\ 2003b, \apj, 598, 1017
\bibitem[Draine(1994)]{draine1994} Draine, B.~T., \& Flatau, P.~J. \ 1994, JOSAA, 11, 1491
\bibitem[Draine(2007)]{draine2007} Draine, B.~T., \& Li, A.\ 2007, \apj, 657, 810
\bibitem[Fitzpatrick(1999)]{fitzpatrick1999} Fitzpatrick, E.~L. \ 1999, \pasp, 111, 63
\bibitem[Flaherty(2007)]{flaherty2007} Flaherty, K.~M., Pipher, J.~L., Megeath, S.~T., et al.\ 2007, \apj, 663, 1069
\bibitem[Gao(2009)]{gao2009} Gao, J., Jiang, B.~W., \& Li, A. \ 2009, \apj, 707, 89
\bibitem[Gilmore(1983)]{gilmore1983} Gilmore, G., \& Reid, N. \ 1983, \mnras, 202, 1025
\bibitem[Girardi(2005)]{girardi2005} Girardi, L., Groenewegen, M.~A.~T., Hatziminaoglou, E., \& da Costa,  L. \ 2005, \aap, 436, 895
\bibitem[Hensley(2016)]{hensley2016} Hensley, B.~S., Draine, B.~T., \& Meisner, A.~M. \ 2016 \apj, 827, 45
\bibitem[Henyey(1941)]{henyey1941} Henyey, L.~G., \& Greenstein, J.~L.\ 1941, \apj, 93, 70
\bibitem[Ienaka(2013)]{ienaka2013} Ienaka, N., Kawara, K., Matsuoka, Y., et al.\ 2013, \apj, 767, 80
\bibitem[Indebetouw(2005)]{indebetouw2005} Indebedouw, R., Mathis, J.~S., Babler, B.~L., et al.\ 2005, \apj, 619, 931
\bibitem[Jiang(2006)]{jiang2006} Jiang, B.~W., Gao, J., Omont, A., Schuller, F., \& Simon, G. \ 2006, \aap, 446, 551
\bibitem[Jura(1979)]{jura1979} Jura, M. \ 1979, \apj, 227, 798
\bibitem[Kawara(2017)]{kawara2017} Kawara, K., Matsuoka, Y., Sano, K., et al.\ 2017, \pasj, 69, 31
\bibitem[Lagache(2000)]{lagache2000} Lagache, G., Haffner, L.~M., Reynolds, R.~J., \& Tufte, S.~L.\ 2000, \aap, 354, 247
\bibitem[Lagache(2007)]{lagache2007} Lagache, G., Bavouzet, N., Fernandez-Conde, N., et al. \ 2007, \apjl, 665, L89
\bibitem[Lehtinen(1996)]{lehtinen1996} Lehtinen, K., \& Mattila, K.\ 1996, \aap, 309, 570
\bibitem[Lutz(1999)]{lutz1999} Lutz, D., \ 1999, in The Universe as Seen by ISO, Vol. 427, ed. P. Cox \& M. Kessler (Noordwijk: ESA Special Publ.), 623
\bibitem[Lynga(1982)]{lynga1982} Lyng\aa, G. \ 1982, \aap, 109, 213
\bibitem[Malhotra(1995)]{malhotra1995} Malhotra, S. \ 1995, \apj, 448, 138
\bibitem[Mathis(1983)]{mathis1983} Mathis, J.~S., Mezger, P.~G., \& Panagia, N.\ 1983, \aap, 128, 212
\bibitem[Matsumoto(2011)]{matsumoto2011} Matsumoto, T., Seo, H.~J., Jeong, W.-S., et al.\ 2011, \apj, 742, 124 
\bibitem[Matsuoka(2011)]{matsuoka2011} Matsuoka, Y., Ienaka, N., Kawara, K., \& Oyabu, S.\ 2011, \apj, 736, 119
\bibitem[Matsuura(2011)]{matsuura2011} Matsuura, S., Shirahata, M., Kawada, M., et al. \ 2011, \apj, 737, 2 
\bibitem[Meisner(2015)]{meisner2015} Meisner, A.~M., \& Finkbeiner, D.~P. \ 2015, \apj, 798, 88
\bibitem[Mendez(1998)]{mendez1998} M\'endez, R.~A., \& van Altena, W.~F. \ 1998, \aap, 330, 910
\bibitem[Mie(1908)]{mie1908} Mie, G. \ 1908, Anp, 330, 377
\bibitem[Mishchenko(1996)]{mishchenko1996} Mishchenko, M.~I., Travis, L.~D., \& Mackowski, D.~W. \ 1996, JQSRT, 55, 535
\bibitem[Murthy(2010)]{murthy2010} Murthy, J., Henry, R.~C., \& Sujatha, N.~V. \ 2010, \apj, 724, 1389
\bibitem[Murthy(2016)]{murthy2016} Murthy, J. \ 2016, \mnras, 459, 1710
\bibitem[Nakanishi(2003)]{nakanishi2003} Nakanishi, H., \& Sofue, Y.\ 2003, \pasj, 55, 191  
\bibitem[Nishiyama(2009)]{nishiyama2009} Nishiyama, S., Tamura, M., Hatano, H., et al. \ 2009, \apj, 696, 1407
\bibitem[Planck(2014)]{planck2014} Planck Collaboration XI. \ 2014, \aap, 571, A11
\bibitem[Planck(2016)]{planck2016} Planck Collaboration XXIX. \ 2016, \aap, 586, A132
\bibitem[Purcell(1973)]{purcell1973} Purcell, E.~M., \& Pennypacker, C.~R.\ 1973, \apj, 186, 705 
\bibitem[Rana(1992)]{rana1992} Rana, N.~C., \& Basu, S. \ 1992, \aap, 265, 499
\bibitem[Sano(2015)]{sano2015} Sano, K., Kawara, K., Matsuura, S., et al.\ 2015, \apj, 811, 77 
\bibitem[Sano(2016a)]{sano2016a} Sano, K., Kawara, K., Matsuura, S., et al.\ 2016a, \apj, 818, 72 
\bibitem[Sano(2016b)]{sano2016b} Sano, K., Matsuura, S., Tsumura, K., et al.\ 2016b, \apjl, 821, L11 (Paper I) 
\bibitem[Schlegel(1998)]{schlegel1998} Schlegel, D.~J., Finkbeiner, D.~P., \& Davis, M.\ 1998, \apj, 500, 525 
\bibitem[Sellgren(1992)]{sellgren1992} Sellgren, K., Werner, M.~W., \& Dinerstein, H.~L. \ 1992, \apj, 400, 238
\bibitem[Sellgren(1996)]{sellgren1996} Sellgren, K., Werner, M.~W., \& Allamandola, L.~J. \ 1996, \apjs, 102, 369 
\bibitem[Tazaki(2016)]{tazaki2016} Tazaki, R., Tanaka, H., Okuzumi, S., Kataoka, A., \& Nomura, H. \ 2016, \apj, 823, 70
\bibitem[Tsumura(2013b)]{tsumura2013b} Tsumura, K., Matsumoto, T., Matsuura, S., et al.\ 2013, \pasj, 65, 120
\bibitem[Wang(2013)]{wang2013} Wang, S., Gao, J., Jiang, B.~W., Li, A., \& Chen, Y. \ 2013, \apj, 773, 30
\bibitem[Wang(2015)]{wang2015} Wang, S., Li, A., \& Jiang, B.~W. \ 2015, \apj, 811, 38
\bibitem[Weingartner \& Draine(2001)]{weingartner2001} Weingartner, J.~C., \& Draine, B.~T.\ 2001, \apj, 548, 296 
\bibitem[Witt(1977)]{witt1977} Witt, A.~N. \ 1977, \apjs, 35, 1
\bibitem[Witt(1994)]{witt1994} Witt, A.~N., Lindell, R.~S., Block, D.~L., \& Evans, R. \ 1994, \apj, 427, 227
\bibitem[Witt(2008)]{witt2008} Witt, A.~N., Mandel, S., Sell, P.~H., Dixon, T., \& Vijh, U.~P. \ 2008, \apj, 679, 497
\bibitem[Witt(1996)]{witt1996} Witt, A.~N., \& Gordon, K.~D. \ 1996, \apj, 463, 681
\bibitem[Yahata(2007)]{yahata2007} Yahata, K., Yonehara, A., Suto, Y., et al. \ 2007, \pasj, 59, 295
\bibitem[Yue(2013)]{yue2013} Yue, B., Ferrara, A., Salvaterra, R., Xu, Y., \& Chen, X. \ 2013, \mnras, 433, 1556
\bibitem[Zemcov(2014)]{zemcov2014} Zemcov, M., Smidt, J., Arai, T., et al.\ 2014, Science, 346, 732
\bibitem[Zubko(2004)]{zubko2004} Zubko, V., Dwek, E., \& Arendt, R.~G.\ 2004, \apjs, 152, 211 
\end{thebibliography}
\end{document}